\def\asca{{\it ASCA\/}}
\def\chandra{{\it Chandra\/}}
\def\conx{{\it Constellation-X\/}}
\def\genx{{\it Generation-X\/}}
\def\xeus{{\it XEUS\/}}
\def\xmm{{\it XMM-Newton\/}}

\def\aox{$\alpha_{\rm ox}$}

\def\sdssfulla{SDSSp~J083643.85+005453.3}
\def\sdssa{SDSS~0836+0054}

\def\sdssfullb{SDSSp~J130608.26+035626.3}
\def\sdssb{SDSS~1306+0356}

\def\sdssfullc{SDSSp~J103027.10+052455.0}
\def\sdssc{SDSS~1030+0524}

\def\sdssfulld{SDSSp~J104433.04--012502.2}
\def\sdssd{SDSS~1044--0125}

\def\ltsima{$\; \buildrel < \over \sim \;$}
\def\simlt{\lower.5ex\hbox{\ltsima}}
\def\gtsima{$\; \buildrel > \over \sim \;$}
\def\simgt{\lower.5ex\hbox{\gtsima}}

\documentclass[preprint]{aastex}
\usepackage{psfig}

\begin{document}


\title{Exploratory Chandra Observations of the Three Highest Redshift Quasars Known} 

\author{
William N.~Brandt,\altaffilmark{1} 
Donald P.~Schneider,\altaffilmark{1}  
Xiaohui Fan,\altaffilmark{2}  
Michael A.~Strauss,\altaffilmark{3}  
James E.~Gunn,\altaffilmark{3}
Gordon T.~Richards,\altaffilmark{1} 
Scott F.~Anderson,\altaffilmark{4}  
Daniel E.~Vanden~Berk,\altaffilmark{5}  
Neta~A.~Bahcall,\altaffilmark{3}  
J.~Brinkmann,\altaffilmark{6}  
Robert Brunner,\altaffilmark{7}  
Bing Chen,\altaffilmark{8}  
G.S.~Hennessy,\altaffilmark{9}
Donald Q.~Lamb,\altaffilmark{10}
Wolfgang Voges,\altaffilmark{11}  
and
Donald G.~York\altaffilmark{10}}
\altaffiltext{1}{Department of Astronomy \& Astrophysics, 525 Davey
Laboratory, The Pennsylvania State University, University Park, PA 16802}
\altaffiltext{2}{Institute for Advanced Study, Olden Lane, Princeton, NJ 08540}
\altaffiltext{3}{Princeton University Observatory, Princeton, NJ 08544}
\altaffiltext{4}{University of Washington, Department of Astronomy, Box 351580, Seattle, WA 98195}
\altaffiltext{5}{Fermi National Accelerator Laboratory, P.O. Box 500, Batavia, IL 60510}
\altaffiltext{6}{Apache Point Observatory, P.O. Box 59, Sunspot, NM 88349-0059}
\altaffiltext{7}{Astronomy Department, California Institute of Technology, Pasadena, CA 91125}
\altaffiltext{8}{European Space Agency-Vilspa, Villafranca del Castillo, Apartado 50727, 28080 Madrid, Spain}
\altaffiltext{9}{US Naval Observatory, 3450 Massachusetts Avenue NW, Washington, DC 20392-5420}
\altaffiltext{10}{University of Chicago, Astronomy \& Astrophysics Center, 5640 S. Ellis Ave., Chicago, IL 60637}
\altaffiltext{11}{Max-Planck-Institut f\"ur Extraterrestrische Physik, Postfach 1603, Garching, D-85740, Germany}


\begin{abstract}
We report on exploratory \chandra\ observations of the three highest
redshift quasars known ($z$ = 5.82, 5.99, and 6.28), all found in the Sloan Digital
Sky Survey.  These data, combined with a previous \xmm\ observation of a
$z = 5.74$ quasar, form a complete set of color-selected, $z > 5.7$ quasars.
X-ray emission is detected from all of the quasars at levels
that indicate that the X-ray to optical flux ratios of $z\approx 6$
optically selected quasars are similar to those of lower redshift quasars.  
The observations demonstrate that it will be feasible to obtain quality 
X-ray spectra of $z \approx 6$ quasars with current and future X-ray missions.

\end{abstract}




\section{Introduction}

One of the main themes in astronomy over the coming decades will be the study of 
the first generation of objects to form in the Universe. Due to their large 
luminosities, quasars are among the most accessible of these, particularly at 
high energies where stars produce little emission. X-ray studies of high-redshift 
quasars reveal the conditions in the immediate vicinity of their supermassive 
black holes. Measurements of the X-ray continuum's shape, amplitude relative
to longer wavelength radiation, and variability can provide information about 
the inner accretion disk and its corona, and thus ultimately about how the black 
hole is fed. The penetrating nature of X-rays allows even highly obscured black 
holes to be probed. While some very high-redshift quasars have been
discovered via their X-ray emission (e.g., Henry et~al. 1994; Zickgraf et~al. 1997; 
Schneider et~al.~1998; Silverman et~al.~2002; A.J. Barger et~al., in preparation), 
the vast majority of distant quasars were identified by optical observations.

Recently, the Sloan Digital Sky Survey (SDSS; York et~al. 2000)
discovered the three highest redshift quasars known to date: \sdssfulla\ at $z=5.82$, 
\sdssfullc\ at $z=6.28$, and \sdssfullb\ at $z=5.99$ (Fan et~al. 2001; 
hereafter these quasars will be referred to through their abbreviated names). 
The SDSS uses a CCD camera (Gunn et~al. 1998) on a dedicated 2.5~m telescope at 
Apache Point Observatory in New Mexico to obtain images in five broad optical 
bands ($u$, $g$, $r$, $i$, $z$; Fukugita et~al. 1996; Stoughton et~al. 2002);
the images of all three of these quasars have extremely red colors in the SDSS 
system. All three quasars are luminous and, by the Eddington argument
(see \S1.2 of Frank, King, \& Raine 1992; Fan et~al. 2001), have central black 
holes with masses of several times $10^9$~$M_\odot$ that formed within a billion 
years of the Big Bang.

We have begun a project to determine the X-ray properties of the highest 
redshift quasars utilizing both the new generation of X-ray observatories
and archival data (Kaspi et~al. 2000; Brandt et~al. 2001; Vignali et~al. 2001). 
Thus far we have mostly been making short, exploratory observations designed
to define the basic X-ray properties of these quasars and to identify
good candidates for future X-ray spectroscopy.
Aside from addressing 
the scientific issues mentioned above, this project is also laying 
groundwork for future X-ray observatories focused on studying the high-redshift
X-ray Universe (e.g., \conx, \xeus, and \genx). After the discoveries of 
\sdssa, \sdssc, and \sdssb, we requested Director's Discretionary Time in 
2001 August for exploratory \chandra\ observations of these quasars. 
This request was approved, and the observations were performed in 
2002 January. Here we present the results of the observations. 

We adopt $H_0=65$~km~s$^{-1}$ Mpc$^{-1}$, 
$\Omega_{\rm M}=1/3$, and 
$\Omega_{\Lambda}=2/3$
throughout. 


\section{Observations and data analysis}

The basic properties of the observed quasars, along with the observation dates
and exposure times, are given in Table~1. We have also included in Table~1 the 
$z=5.74$ Broad Absorption Line (BAL) quasar 
\sdssfulld\ (Fan et~al. 2000; 
Maiolino et~al.~2001; Goodrich et~al. 2001; hereafter \sdssd); 
this quasar has not been observed by \chandra, but it has been detected by \xmm\ 
(Brandt et~al. 2001). We have included \sdssd\ in this paper
because it and the 
other three quasars form a complete $z\simgt 5.7$ color-selected sample  
down to a $z$ magnitude of $\approx 20$ over about 1500~deg$^{2}$
(Fan et~al. 2001). \sdssa\ has a 20~cm radio detection of 1.2~mJy
in the FIRST survey (Becker, White, \& Helfand 1995) and is 
radio intermediate ($R=8.5$; see Table~1 for the definition of $R$). 
The other quasars lack radio detections and could be either radio 
quiet or radio intermediate (see Table~1 for their $R$ parameters). 

All targets were observed at the aimpoint of the S3 back-illuminated CCD in the 
\chandra\ Advanced CCD Imaging Spectrometer (ACIS; G.P. Garmire et~al., in 
preparation). Given previous X-ray studies of \hbox{$z>4$} quasars and the superb source
detection capability of \chandra, these quasars were expected to be detectable 
with relatively short 6--8~ks \chandra\ observations. Faint mode was used for 
the event telemetry format, and \asca\ grades 0, 2, 3, 4 and 6 were used in 
all analysis; this grade set choice is a standard one that, in general, 
optimizes signal-to-noise ratio.\footnote{See \S6.3 of the \chandra\ Proposers' 
Observatory Guide at http://asc.harvard.edu/udocs/docs/docs.html for a discussion 
of grades.} We have searched for background flares that occurred during
the observations and found none. We have inspected the photon arrival times
for the X-ray sources described below, and there is no evidence for transient 
spurious phenomena affecting the data. 

We created images around the quasar positions in each of the four standard 
bands defined in Table~2, and we show adaptively smoothed full-band 
images for each of the quasars in Figure~1. Note that in the rest frame
we are probing the $\approx$~2--60~keV emission from these quasars. Source 
detection was performed with {\sc wavdetect} (Freeman et~al. 2002). For
each image, we calculated wavelet transforms (using a Mexican hat kernel)
with wavelet scale sizes from 1--4 pixels. Those peaks whose probability of
being false were less than the threshold of $10^{-5}$ were taken as real. 
All of the quasars were detected in at least two of the standard bands 
(see Table~2). The photometry in Table~2 was performed using circular 
apertures with radii of $2^{\prime\prime}$; errors on the photometry due
to background subtraction are negligible. 
In all cases, the full-band X-ray centroid positions lie within 
$1^{\prime\prime}$ of the precise optical positions of the quasars; this 
is within plausible errors considering the limited number of counts.  
Given the positional coincidence, the probability of an unrelated, 
confusing counterpart for even our X-ray faintest quasar (\sdssc) is 
only $\sim 1\times 10^{-4}$.  

Since our observations do not have sufficient counts for spectral fitting, we 
have calculated the quasars' fluxes and luminosities adopting a nominal
power-law
model with a photon index of $\Gamma=2$ (at energy $E$ the photon density
$N(E)\propto E^{-\Gamma}$; e.g., Reeves \& Turner 2000) 
and the Galactic absorption column densities in Table~1. This spectral
model is consistent with the observed hardness ratios of our targets, although
the constraints on the hardness ratios are poor due to the limited numbers of 
counts. Using the \chandra\ X-ray Center Portable Interactive Multi-Mission
Simulator ({\sc pimms}; Mukai 2001), we find the absorption-corrected
0.5--2.0~keV fluxes given in Table~2. In Table~2 we also present the 
derived luminosities in the 3.5--14.0~keV rest-frame band; this 
rest-frame band is well matched to the observed 0.5--2.0~keV band
for our objects. Note that for a $\Gamma=2$ power law the 
rest-frame 3.5--14.0~keV and 0.5--2.0~keV luminosities are 
the same. 

We have calculated \aox, the slope of a nominal power law between 2500~\AA\ and 2~keV
in the rest frame [$\alpha_{\rm ox}=0.384\log (f_{\rm 2~keV}/f_{2500~\mbox{\footnotesize \AA}}$) 
where $f_{\rm 2~keV}$ is the flux density at 2~keV and
$f_{2500~\mbox{\footnotesize \AA}}$ is the flux density at 2500~\AA], for 
each of our targets (see Table~2). We again adopt $\Gamma=2$ for the X-ray 
continuum, and we use an optical power-law slope of $\alpha_{\rm o}=-0.5$ 
($f_\nu\propto \nu^{\alpha_{\rm o}}$; e.g., Schneider et~al. 2001; 
Vanden~Berk et~al. 2001) to estimate the flux density at 2500~\AA\ in 
the rest frame. We have used the observed soft-band fluxes to find
$f_{\rm 2~keV}$ in the \aox\ calculations, so the derived \aox\ values 
are actually based on the relative amount of X-ray 
emission in the $\approx$~3.5--14.0~keV rest-frame band; 2~keV in the 
rest frame corresponds to $\approx 0.3$~keV in the observed frame for 
these quasars, and this energy is poorly sampled by ACIS for these faint 
sources. 


\section{Discussion}

These three short \chandra\ observations provide the highest redshift 
X-ray detections to date, demonstrating the power of \chandra\ for 
probing the high-redshift X-ray Universe efficiently. Figure~2 shows a redshift
histogram of the known \hbox{$z > 4$} quasars; the X-ray detections are
indicated in the figure.\footnote{The data used to construct this figure are
available from http://www.astro.caltech.edu/$\sim$george/z4.qsos and
http://www.astro.psu.edu/users/niel/papers/highz-xray-detected.dat.} 
Together with \sdssd, the highest redshift X-ray detection obtained 
previously, the quasars studied here form a small
but complete $z\simgt 5.7$ color-selected sample that should be
representative of the luminous, optically selected quasar population at 
$z\approx 6$ (we recognize that optically selected $z\approx 6$ quasars
may not be representative of the entire $z\approx 6$ quasar population, 
but these quasars are the only ones available for study at present). 

In Figure~3 we plot the Galactic absorption-corrected 0.5--2.0~keV flux versus $AB_{1450}$ 
magnitude for the quasars in Table~1 as well as for other $z>4$ quasars 
(e.g., Kaspi et~al. 2000; Vignali et~al. 2001; Silverman et~al. 2002).
While the BAL quasar \sdssd\ is notably X-ray weak (probably due to 
intrinsic X-ray absorption; see Brandt et~al. 2001), the $z\approx 6$ 
quasars observed by \chandra\ appear to lie within the locus of points 
for other $z>4$ quasars. The spectral region around the C~{\sc iv} line 
at rest-frame 1549~\AA\ has been observed in both \sdssc\ and \sdssb; 
neither shows evidence for BALs (Fan et~al. 2001; Pentericci et~al. 2002). 
The integrated column density of the intergalactic medium to these quasars, 
including both ionized and neutral material, is almost certainly too small 
to produce significant X-ray absorption (e.g., Weinberg et~al. 1997;  
Miralda-Escud\'e 2000). This is especially true given the intergalactic 
medium's low metallicity.  

Figure~4 shows that there is no strong evolution in \aox\ for optically 
selected, radio-quiet quasars (RQQs) out to $z\approx 6$ (despite the
strong changes in quasar number density over the redshift range shown
in this figure; see \S5.3 of Fan et~al. 2001 and references therein). 
The central X-ray power sources of quasars do not appear to evolve strongly 
out to this redshift, and there is no indication that strong intrinsic 
obscuration of the X-ray emission generally occurs at $z\approx 6$. 
This result bodes well for attempts to detect the first massive black holes to form in 
the Universe ($z\approx$~8--20) with deep X-ray surveys; our data 
suggest that these objects are likely to be luminous X-ray emitters. 
Furthermore, this result helps to validate the bolometric correction 
factor adopted by Fan et~al. (2001) when estimating the black hole 
masses of these objects via the Eddington argument. 
Some studies have found evidence that \aox\ depends upon quasar
luminosity, with more luminous quasars generally having larger negative
values of \aox\ (e.g., Green et~al. 1995 and references therein). 
The $z\approx 6$ quasars under study here have comparable luminosities
at 2500~\AA\ to those of the typical $z\approx$~4--5 quasars studied by
Vignali et~al. (2001; see their Figure~5 noting the different cosmology). 
Therefore, we do not expect luminosity effects upon \aox\ to affect
our conclusions materially (see \S4.2 of Vignali et~al. 2001 for further
discussion). 

The observations presented here are consistent with the statement that the
majority of optically selected quasars at redshifts of near six possess
similar X-ray properties to those of their low-redshift counterparts.
X-ray observations have now reached what appears to be the redshift regime
of the reionization of the intergalactic medium (Becker et~al.~2001;
Djorgovski et~al. 2001). The \chandra\ observations also indicate that X-ray
spectroscopy with \xmm\ will be feasible for a significant fraction of 
$z\approx 6$ quasars (with $\approx 100$~ks exposures), and high-quality 
X-ray spectroscopy of this class of objects will be possible 
with \conx, \xeus, and \genx .



\acknowledgments
This work would not have been possible without the enormous efforts 
of the entire \chandra\ team. 
We thank H.D.~Tananbaum for kindly allocating the time for these observations, and 
we thank S.N.~Virani for help with observation planning. 
We thank S.C.~Gallagher, S.~Kaspi, C.~Vignali, and an anonymous referee
for helpful discussions. 
We gratefully acknowledge the financial support of 
NASA LTSA grant NAG5-8107 (WNB),
NSF grant AST-9900703 (DPS, GTR), and  
NSF grant AST-0071091 (MAS). 

The Sloan Digital Sky Survey (SDSS) is a joint project of The
University of Chicago, Fermilab, the Institute for Advanced Study, the
Japan Participation Group, The Johns Hopkins University, the
Max-Planck-Institute for Astronomy (MPIA), the Max-Planck-Institute for
Astrophysics (MPA), New Mexico State University, Princeton University,
the United States Naval Observatory, and the University of Washington.
Apache Point Observatory, site of the SDSS telescopes, is operated by
the Astrophysical Research Consortium (ARC).
Funding for the project has been provided by the Alfred P. Sloan
Foundation, the SDSS member institutions, the National Aeronautics and
Space Administration, the National Science Foundation, the
U.S. Department of Energy, the Japanese Monbukagakusho, and the Max
Planck Society. The SDSS Web site is http://www.sdss.org/.


\clearpage


\begin{deluxetable}{lcccccccc}
\tabletypesize{\scriptsize}
\tablewidth{0pt}
\tablecaption {Basic Quasar Properties and Observation Log}
\scriptsize
\tablehead{
\colhead{}                     &
\colhead{}                     & 
\colhead{}                     & 
\colhead{}                     &
\colhead{}                     &
\colhead{}                     &
\colhead{Galactic $N_{\rm H}$} &
\colhead{Obs.}                 &
\colhead{Exp.}                 \\
\colhead{Name}                                        &
\colhead{$z$}                                         &
\colhead{$AB_{1450}$}                                 &
\colhead{$f_{2500~\mbox{\scriptsize \AA}}^{\rm a}$}   &
\colhead{$M_{1450}$}                                  &
\colhead{$R^{\rm b}$}                                 &           
\colhead{($10^{20}$~cm$^{-2}$)$^{\rm c}$}             &           
\colhead{Date}                                        &           
\colhead{Time (ks)}                          
}
\startdata
\sdssa           & 5.82 & 18.81 & 14.2  & $-27.9$ & 8.5    & 4.4 & 2002 Jan 29 & 5.7 \\  
\sdssc           & 6.28 & 19.66 & 6.5   & $-27.2$ & $<7.5$ & 2.7 & 2002 Jan 29 & 8.0 \\  
\sdssb           & 5.99 & 19.55 & 7.3   & $-27.2$ & $<8.7$ & 2.1 & 2002 Jan 29 & 8.2 \\  
\hline
\sdssd$^{\rm d}$ & 5.74 & 19.21 & 9.8   & $-27.5$ & $<8.5$ & 4.6 & 2000 May 28 & 40.0 \\
\enddata
\tablenotetext{a}{Observed flux density at $\lambda_{\rm obs}=2500(1+z)$~\AA\ in units of
$10^{-28}$~erg~cm$^{-2}$~s$^{-1}$~Hz$^{-1}$. If rest-frame values are desired, divide the
listed numbers by $(1+z)$ following \S3.5.1 of Weedman (1986); this is the ``bandpass''
correction.}
\tablenotetext{b}{Radio-loudness parameter. $R=f_{\rm 6~cm}/f_{4400~\mbox{\scriptsize \AA}}$ 
where $f_{\rm 6~cm}$ is the rest-frame flux density at 6~cm and
$f_{4400~\mbox{\scriptsize \AA}}$ is the rest-frame flux density at 4400~\AA\
(e.g., Kellermann et~al. 1989). $f_{\rm 6~cm}$ has been calculated using data 
from FIRST (Becker et~al. 1995), and $f_{4400~\mbox{\scriptsize \AA}}$ has been 
calculated using data from Fan et~al. (2001).}
\tablenotetext{c}{From Stark et~al. (1992).}
\tablenotetext{d}{This quasar was not observed by \chandra, but it has 
been detected by \xmm\ (Brandt et~al. 2001). We have included information on 
it here because it and the other three quasars form a complete color-selected 
sample at $z\simgt 5.7$. The observation date and exposure time apply to 
the \xmm\ observation.}
\end{deluxetable}

\clearpage


\begin{deluxetable}{lcccccccc}
\tabletypesize{\scriptsize}
\tablewidth{0pt}
\tablecaption {X-ray Counts and Properties}
\scriptsize
\tablehead{
\colhead{}                           &
\multicolumn{4}{c}{Counts$^{\rm a}$} &
\colhead{}                           &
\colhead{}                           &
\colhead{}                           \\
\colhead{}                 &
\colhead{Ultrasoft}        &
\colhead{Soft}             &
\colhead{Hard}             &
\colhead{Full}             &
\colhead{}                 &
\colhead{}                 &
\colhead{}                 &
\colhead{}                 \\
\colhead{Name}                         &
\colhead{0.3--0.5~keV}                 & 
\colhead{0.5--2.0~keV}                 & 
\colhead{2.0--8.0~keV}                 &
\colhead{0.5--8.0~keV}                 &
\colhead{$F_{0.5-2.0}^{\rm b}$}        &           
$f_{\rm 2~keV}^{\rm c}$                &
\colhead{$L_{3.5-14.0}^{\rm d}$}       &           
\colhead{$\alpha_{\rm ox}^{\rm e}$}                             
}
\startdata
\sdssa           & $<6.4$  & $16.9^{+5.2}_{-4.1}$  & $3.8^{+3.1}_{-1.9}$  & $20.7^{+5.6}_{-4.5}$ & 10.29 & 10.5 & 45.6 & $-1.58^{+0.05}_{-0.05}$  \\ 
\sdssc           & $<4.8$  & $5.9^{+3.6}_{-2.4}$   & $<3.0$               & $5.8^{+3.6}_{-2.3}$  & 2.43  & 2.6  & 45.1 & $-1.68^{+0.08}_{-0.09}$  \\ 
\sdssb           & $<4.8$  & $11.8^{+4.5}_{-3.4}$  & $4.9^{+3.4}_{-2.1}$  & $16.8^{+5.2}_{-4.1}$ & 4.63  & 4.8  & 45.3 & $-1.60^{+0.05}_{-0.06}$  \\ 
\hline
\sdssd$^{\rm f}$ & $\cdots$ & $\cdots$ & $\cdots$ & $\cdots$                                     & 1.22  & 1.2  & 44.7 & $-1.88^{+0.04}_{-0.05}$  \\  
\enddata
\tablenotetext{a}{Errors on the counts have been calculated following
Gehrels (1986) for $1\sigma$. Upper limits have been calculated following
Kraft, Burrows, \& Nousek (1991) for 95\% confidence.}
\tablenotetext{b}{Galactic absorption-corrected flux in the 0.5--2.0~keV 
observed-frame band in units of $10^{-15}$~erg~cm$^{-2}$~s$^{-1}$.}
\tablenotetext{c}{Observed flux density at $E_{\rm obs}=2/(1+z)$~keV in units of
$10^{-32}$~erg~cm$^{-2}$~s$^{-1}$~Hz$^{-1}$. If rest-frame values are desired, divide the
listed numbers by $(1+z)$ following \S3.5.1 of Weedman (1986); this is the ``bandpass''
correction.}
\tablenotetext{d}{Logarithm of the Galactic absorption-corrected luminosity 
in the 3.5--14.0~keV rest-frame band. This rest-frame band is well 
matched to the observed 0.5--2.0~keV band for our objects.}
\tablenotetext{e}{\aox\ is calculated between 2500~\AA\ and 2~keV. The error bars 
on \aox\ represent the statistical uncertainty associated with the observed number of counts.}
\tablenotetext{f}{The quantities reported here are derived from the \xmm\ observation
using the revised redshift and the same spectral assumptions made for the other
quasars; see Table~1 for further information.}
\end{deluxetable}

\clearpage


\begin{figure}[t!]
\hbox{
\hspace*{0.6 in} \psfig{figure=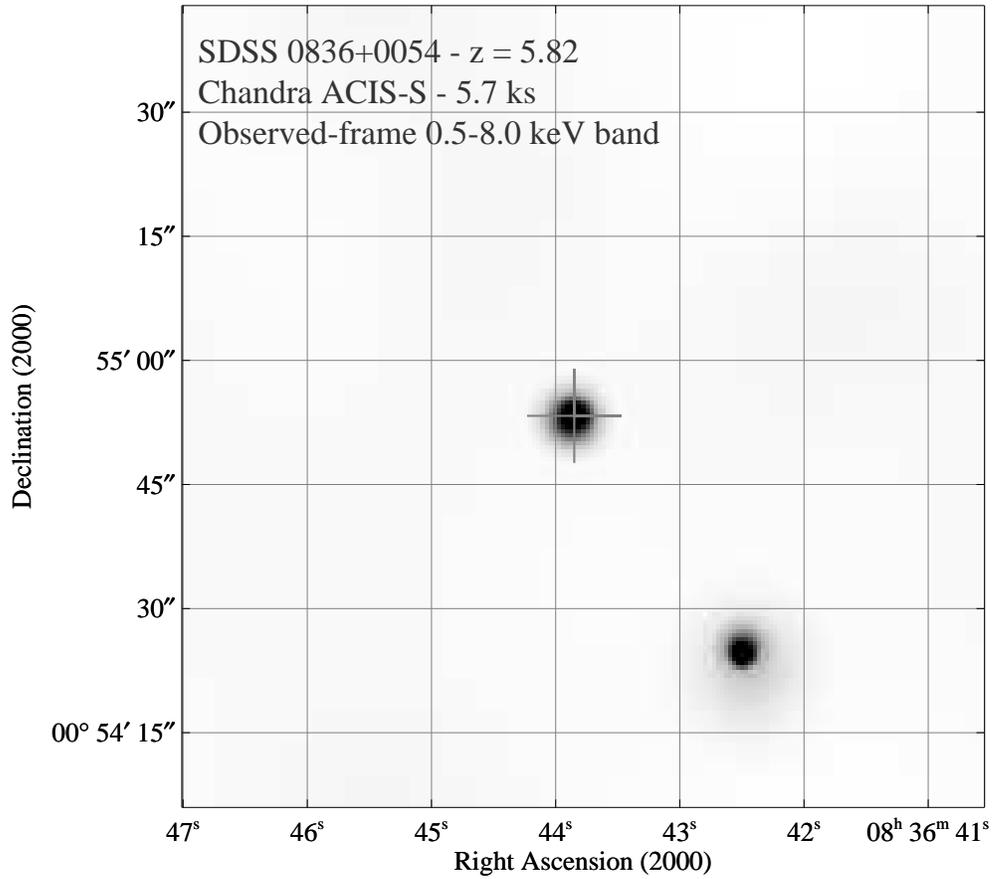,height=5.0truein,width=5.0truein,angle=0}
}
\caption{Full-band ACIS S3 images of the three target quasars. The images 
have been adaptively smoothed at the $2\sigma$ level using the algorithm of 
Ebeling, White, \& Rangarajan (2002). The gray scales are linear. Crosses
mark the optical positions of the quasars (accurate to $\approx 0\farcs 1$ in 
each coordinate).}
\end{figure}

\clearpage

\hbox{
\hspace*{0.6 in} \psfig{figure=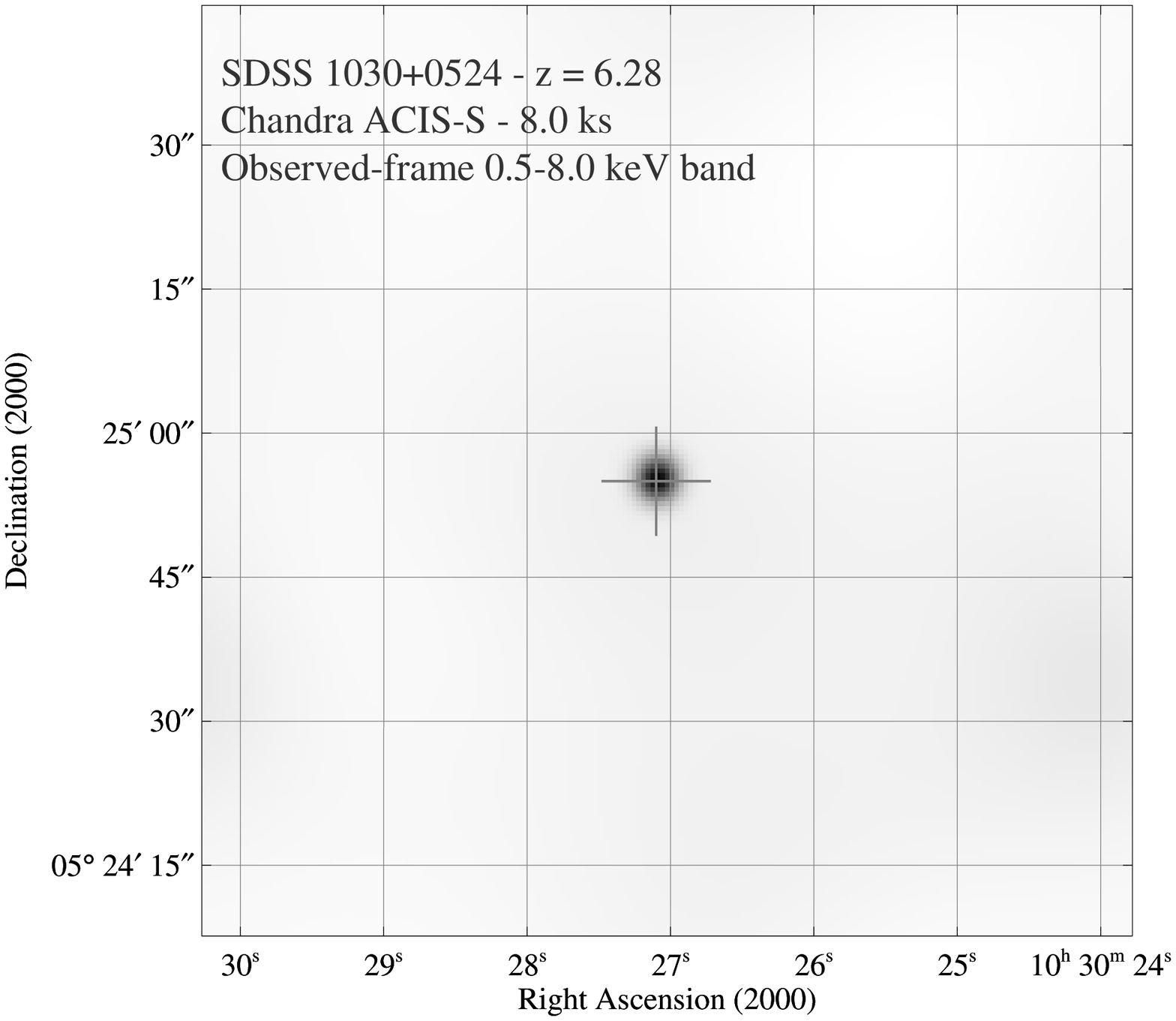,height=5.0truein,width=5.0truein,angle=0}}

\clearpage

\hbox{
\hspace*{0.6 in} \psfig{figure=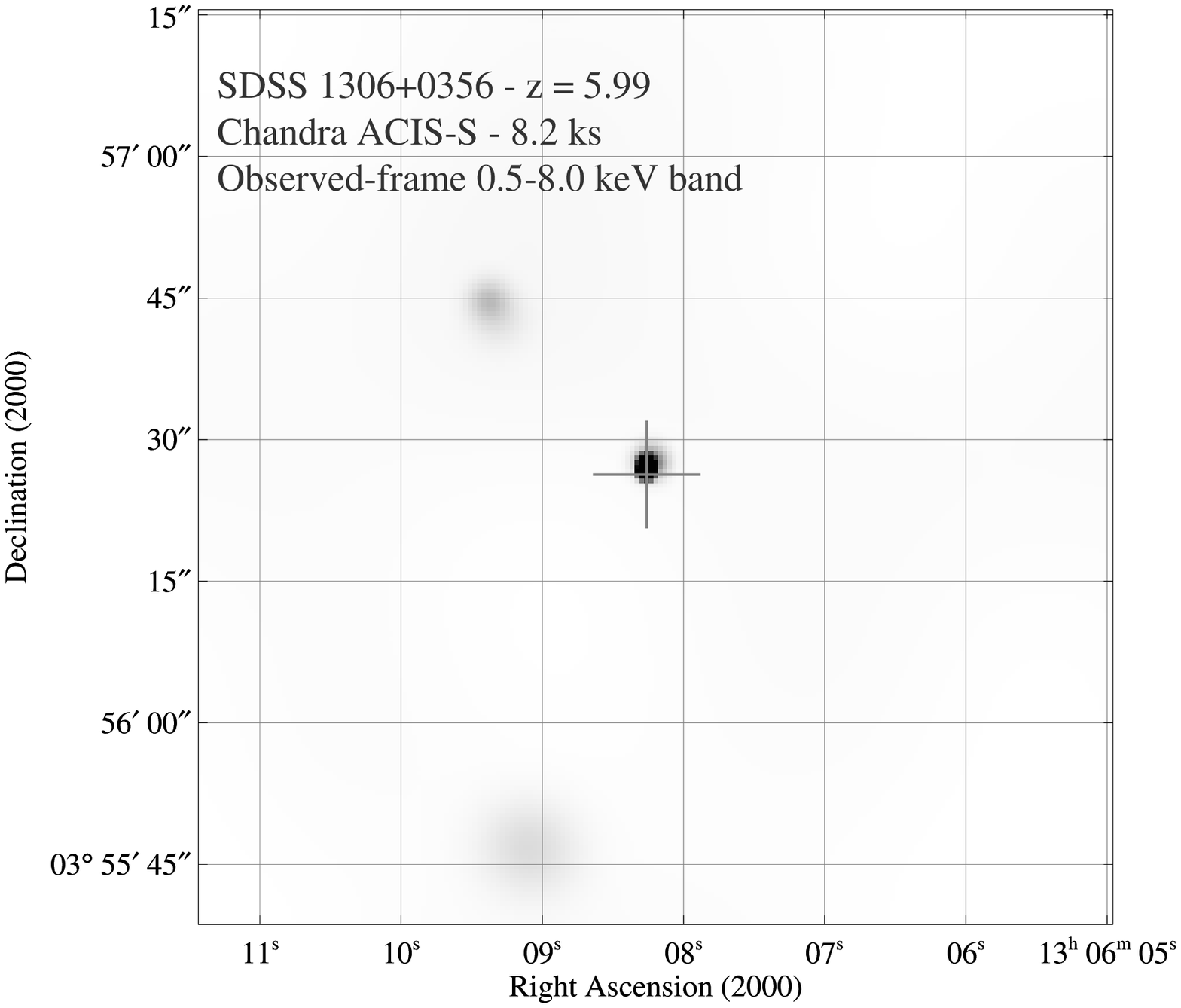,height=5.0truein,width=5.0truein,angle=0}}

\clearpage


\begin{figure}[t!]
\hbox{
\hspace*{0.6 in} \psfig{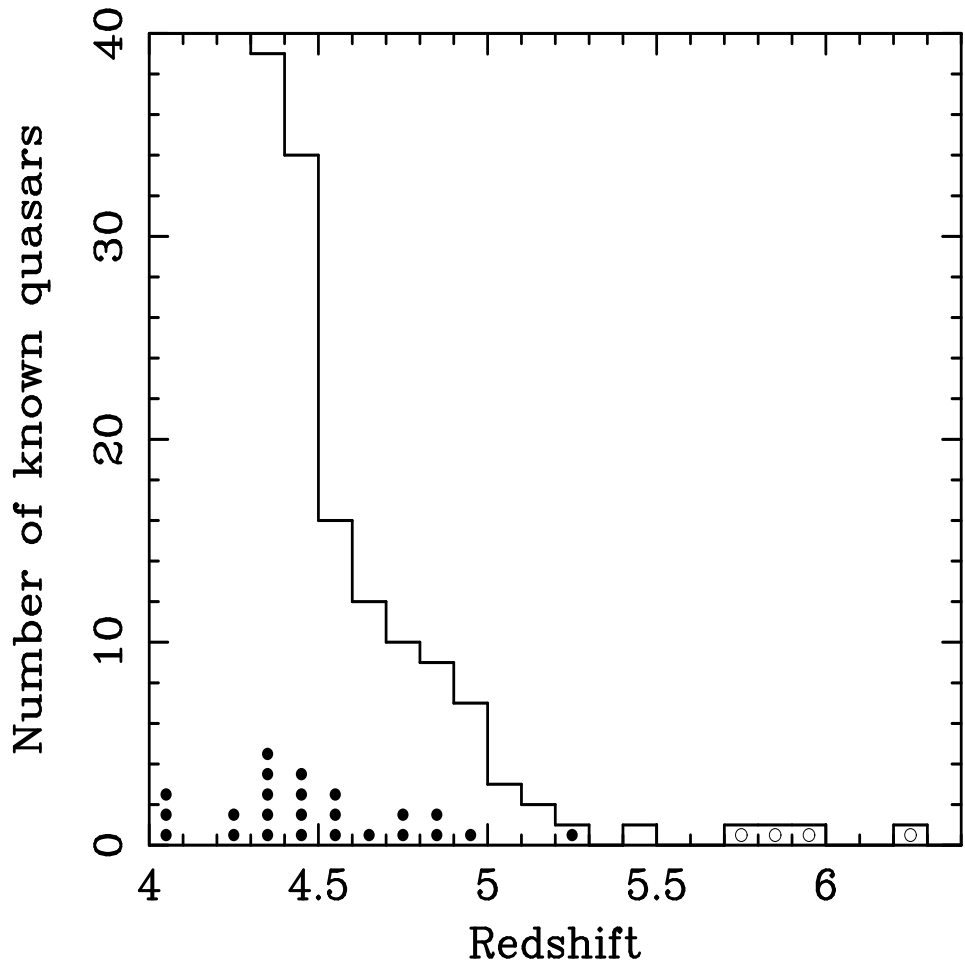}
}
\caption{Redshift distribution of known $z>4$ quasars. The dots 
indicate quasars detected in X-rays previously, and the open circles 
indicate the four quasars under study here including \sdssd. Note that 
most of the known $z\approx$~4--5 quasars do not have X-ray observations 
at present.}
\end{figure}

\clearpage


\begin{figure}[t!]
\hbox{
\hspace*{0.6 in} \psfig{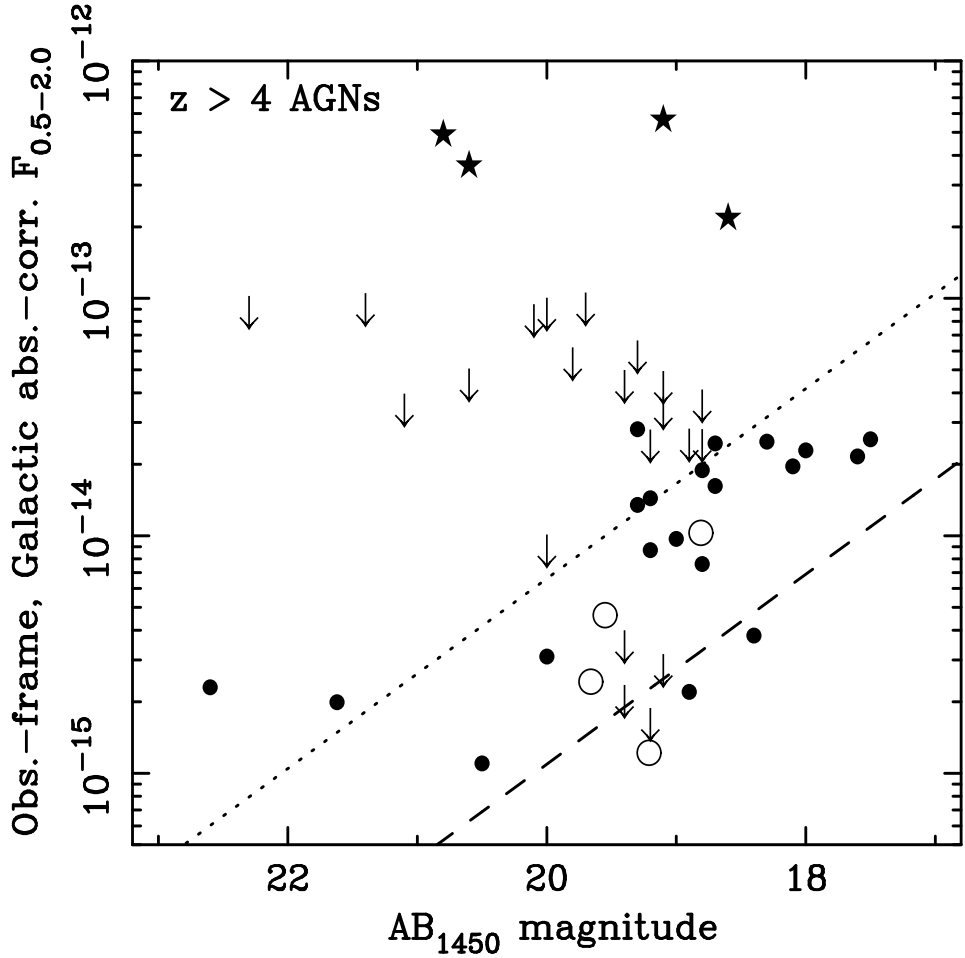}
}
\caption{Observed-frame, Galactic absorption-corrected 0.5--2.0~keV flux versus
$AB_{1450}$ magnitude for $z>4$ AGNs; the units of the ordinate are 
erg~cm$^{-2}$~s$^{-1}$. The dots and arrows show $z>4$ detections 
and upper limits, respectively, from Kaspi et~al. (2000), 
Vignali et~al. (2001), and Silverman et~al. (2002). Blazars 
at $z>4$ are shown as stars. The open 
circles show the quasars studied here including \sdssd. The slanted lines 
show $z=6.0$ loci for $\alpha_{\rm ox}=-1.5$ (dotted) and 
$\alpha_{\rm ox}=-1.8$ (dashed).}
\end{figure}

\clearpage


\begin{figure}[t!]
\hbox{
\psfig{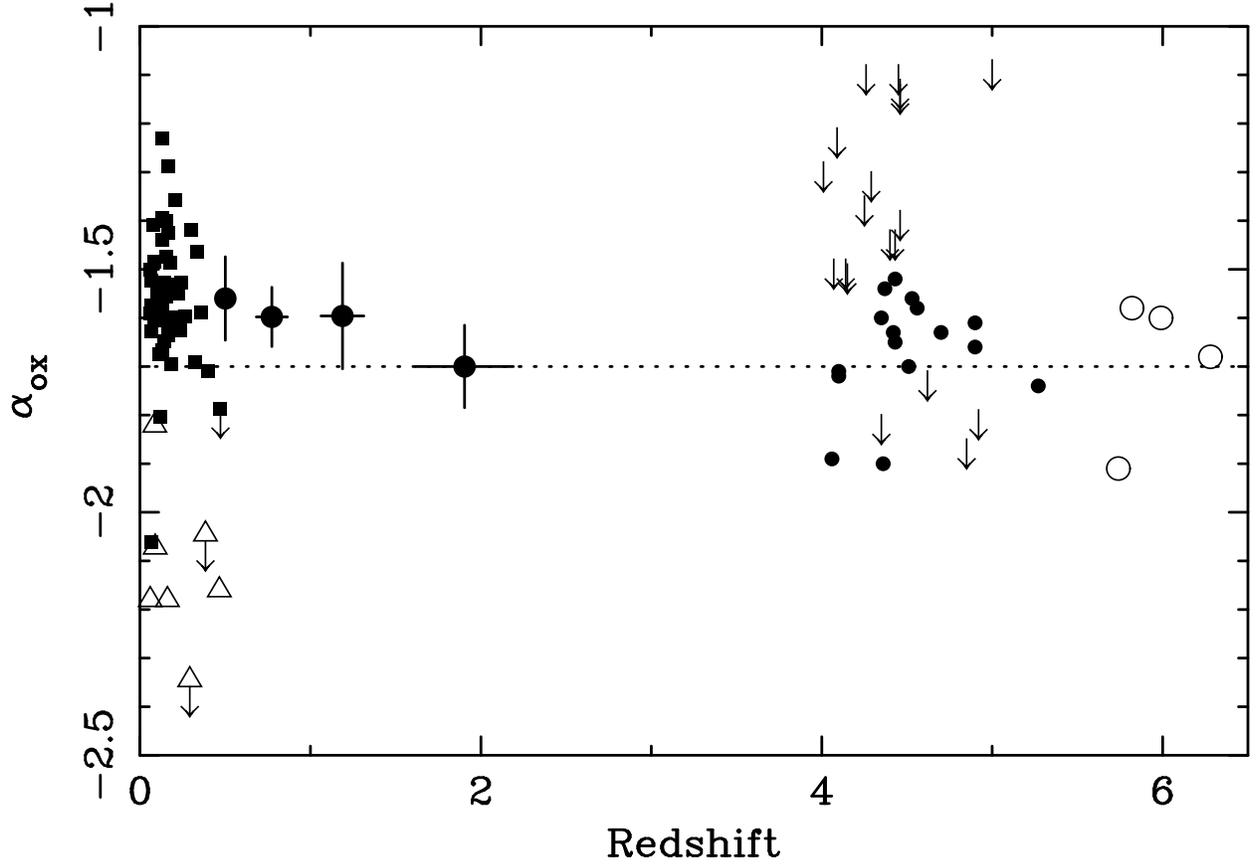}
}
\caption{\aox\ versus redshift for optically selected RQQs. The open triangles are for
seven luminous, absorbed Bright Quasar Survey (BQS; Schmidt \& Green 1983) 
RQQs, and the solid squares are for the other 46 luminous BQS RQQs (from 
Brandt, Laor, \& Wills 2000). The large solid dots with error bars show stacking 
results for Large Bright Quasar Survey (LBQS; Hewett, Foltz, \& Chaffee 1995) 
RQQs from Figure~6d of Green et~al. (1995). The small solid dots and plain arrows
show $z>4$ detections and upper limits, respectively, from 
Kaspi et~al. (2000) and Vignali et~al. (2001). The open circles show the 
quasars studied here including \sdssd. A horizonal line has been drawn 
at $\alpha_{\rm ox}=-1.7$ to guide the eye.}
\end{figure}


\end{document}